# Understanding the Bias Dependence of Low Frequency Noise in Single Layer Graphene FETs

Nikolaos Mavredakis*[a], Ramon Garcia Cortadella[b], Andrea Bonaccini Calia[b], Jose A. Garrido[b] and David Jiménez[a]

This letter investigates the bias-dependent low frequency noise of single layer graphene field-effect transistors. Noise measurements have been conducted with electrolyte-gated graphene transistors covering a wide range of gate and drain bias conditions for different channel lengths. A new analytical model that accounts for the propagation of the local noise sources in the channel to the terminal currents and voltages is proposed in this paper to investigate the noise bias dependence. Carrier number and mobility fluctuations are considered as the main causes of low frequency noise and the way these mechanisms contribute to the bias dependence of the noise is analyzed in this work. Typically, normalized low frequency noise in graphene devices has been usually shown to follow an M-shape dependence versus gate voltage with the minimum near the charge neutrality point (CNP). Our work reveals for the first time the strong correlation between this gate dependence and the residual charge which is relevant in the vicinity of this specific bias point. We discuss how charge inhomogeneity in the graphene channel at higher drain voltages can contribute to low frequency noise; thus, channel regions nearby the source and drain terminals are found to dominate the total noise for gate biases close to the CNP. The excellent agreement between the experimental data and the predictions of the analytical model at all bias conditions confirms that the two fundamental 1/f noise mechanisms, carrier number and mobility fluctuations, must be considered simultaneously to properly understand the low frequency noise in graphene FETs. The proposed analytical compact model can be easily implemented and integrated in circuit simulators, which can be of high importance for graphene based circuits' design.

## Introduction

The outstanding characteristics of graphene such as its high carrier mobility and saturation velocity has attracted significant interest to use this material in future high-performance, high frequency electronics. Although its gapless nature renders it inappropriate for digital circuitry, it can result in a tremendous performance boost in both analog and radio frequency (RF) applications[1-2]. In addition, graphene could also be successfully used in chemical, biological sensors[3-9] as well as in optoelectronic devices[10]. Such applications, though, are extremely prone to Low Frequency Noise (LFN) which can limit the sensitivity of sensors and can also be up-converted to undesired phase noise in voltage controlled oscillators. Furthermore, LFN is a very powerful tool for characterizing the quality and reliability of graphene devices[11-12].

LFN is also referred to as 1/f (flicker) noise when its Power Spectral Density (PSD) is inversely proportional to frequency, which is usually the case in devices with channel lengths typically longer than few hundreds of nanometres. The capture and subsequent emission of charges at border traps near the dielectric interface of oxide semiconductors is the main effect responsible for the generation of LFN[13]. Each carrier that gets trapped causes a Random Telegraph Signal (RTS) in time domain, corresponding to a Lorentzian spectrum determined by a time constant. The high number of such Lorentzians in large devices and the uniform spatial distribution of these traps that results in a uniform distribution of time constants, are responsible for the 1/f behavior of noise. This noise mechanism is called carrier number fluctuation effect ($\Delta N$) and was first proposed by McWhorter[14]. This phenomenon is adequately described by a number of basic LFN models for metal-oxide-semiconductor field-effect transistors (MOSFETs) available in bibliography[15-19]. In addition to carrier number, mobility fluctuation ($\Delta\mu$) is also considered a main contribution to LFN in semiconductor devices and can be generated due to fluctuations in the scattering cross-section of scattering centres. This effect is described by the empirical Hooge formula[20].

In this letter we focus on the effect of LFN on single layer graphene devices (GFETs) and more specifically on long channel solution-gated transistors[21], which are broadly used in biosensing and bioelectronics applications (Fig. 1a). (Details on the fabrication of these devices can be found in Experimental Data section). A map of the 2D/G Raman bands intensity ratio and the average Raman spectrum over the graphene channel are shown in Fig. 1b and 1c respectively. According to the values of the 2D/G band intensity map, a low second nucleation density as well as a relatively good SLG homogeneity can be derived. The D/G ratio in the average spectrum indicates a low density of defects in the graphene channel. Flicker noise which prevails in these transistors, is of high interest because of its unique characteristics[22]. As a semimetal, graphene presents mobility fluctuations which can generate 1/f noise. On the other hand, single-layer graphene (SLG), as a 2D material is extremely prone to trapping effects leading to high amplitude carrier number fluctuations. In fact, a recent study illustrated that LFN can either be dominated by carrier number fluctuations (surface noise) or mobility fluctuations (volume noise) effect depending on the number of Graphene layers[23]; the lower this number the more dominant the surface LFN is. The addition of these two contributions, combined with the unusual transfer characteristics of graphene FETs and the noise originated at the contacts[24] leads to a rather complex dependence of noise on the gate voltage. More specifically, it has been stated that 1/f noise follows a V-shape dependence close to the Dirac or charge neutrality point (CNP) with the minimum of the V-shape at this gate voltage; this behavior can turn into an M-shape in case the gate bias is extended[4, 25-33]. The gate dependence has been shown to strongly depend on the spatial charge inhomogeneity related to the presence of both electron and hole puddles near the CNP[31] and it has been observed in both top-gated[4, 25-30] and back-gated[26-28, 31-33] devices. We will also show that the charge inhomogeneity induced in graphene devices at higher drain voltage values, which is more

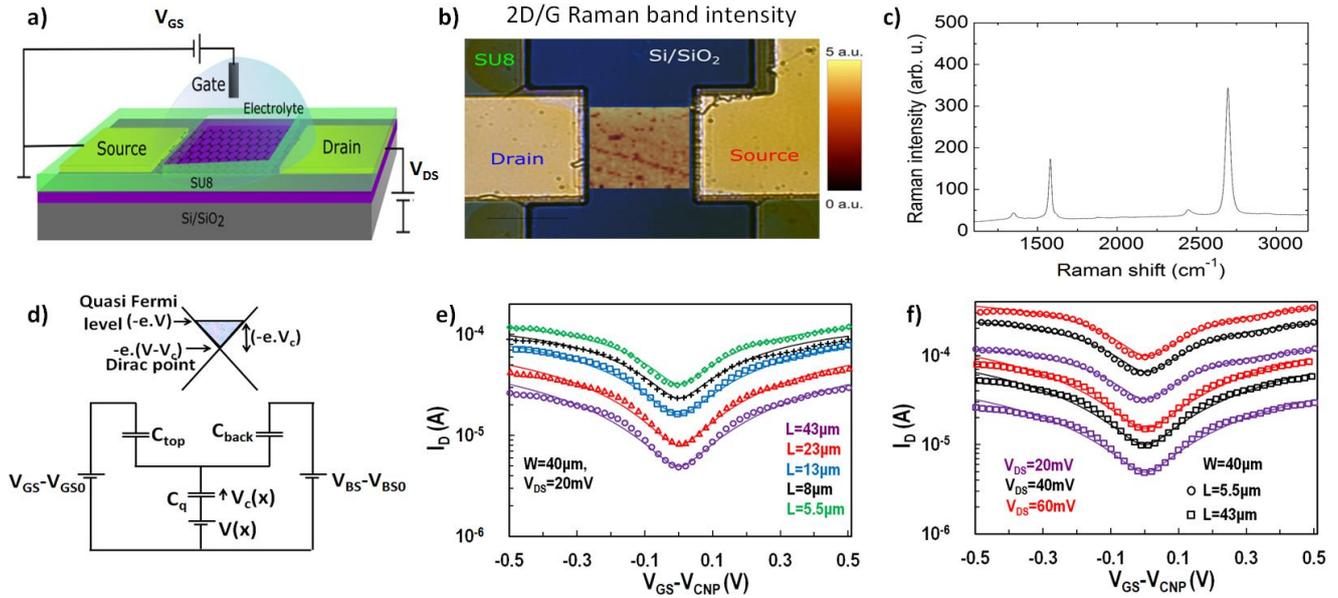

**Fig. 1** a) Schematic of a solution-gated GFET, b) The colour scale indicates the 2D/G Raman band intensity ratio. The colour map overlapped with the optical image of the graphene transistor, represents the local value of the 2D/G ratio measured in the channel area. c) Average Raman spectrum over the whole graphene channel. d) Energy dispersion relation of a single layer GFET (top) and its capacitive network are shown with $C_q$: quantum capacitance, $C_{top}$, $C_{back}$: top and back oxide capacitances, $V_c(x)$: chemical potential, $V(x)$: quasi-Fermi channel potential, $V_{G(B)S}-V_{G(B)S0}$: top and back gate source voltage overdrives. (Back gate is not active in devices under test of (a) but is included in the capacitive network of (d) to support the generalizability of the model) Drain current $I_D$ vs top gate voltage ($V_{GS}-V_{CNP}$) measured in solution -gated GFETs with e) $W=40\ \mu m$ for different channel length values ($L=43,\ 23,\ 13,\ 8,\ 5.5\ \mu m$) at $V_{DS}=20\ mV$ and f) $V_{DS}=20,\ 40,\ 60\ mV$ for $L=43,\ 5.5\ \mu m$. Symbols: experimental data, solid lines: model.

intense at CNP, has a significant effect on the LFN. In case of GFETs on particular substrates such as boron nitride, not only 1/f noise is reduced in comparison to standard SiO₂ substrates but also the M-shape is eliminated or almost disappears[31-32]. It will be shown that the latter occurs in cases where less charge is induced by impurities near CNP, also known as residual charge[34]. Furthermore, flicker noise is shown to be reduced after the effect of electron-beam irradiation[35] while the introduction of graded thickness throughout the graphene channel, with a single layer in the middle and two or more layers close to the contacts, also reduces 1/f noise whereas it still ensures a high mobility[36]. Classical Hooge formula alone cannot predict such M- shape behavior since residual charge does not play a significant role as it will be shown and this can only lead to a Λ-shape gate bias dependence[37]. On the other hand, V and M shapes can be explained in terms of carrier number fluctuations due to charge trapping/detrapping processes[26, 28].

There have been several attempts to model 1/f noise in GFETs considering either carrier number fluctuations[4, 38-42] (ΔN) or mobility fluctuations effects[27] (Δμ), while in some cases both effects have been taken into account simultaneously[28]. Usually noise models are taken from conventional Si devices[4, 28, 41-42] assuming that noise is homogeneously generated over the channel. This assumption is consequence of considering charge to be homogeneously distributed along the channel leading to a carrier number noise which is proportional to the transconductance[4]. In few reports, detailed formulas are derived; however, they are not compact[38-40] and, thus, they cannot be solved analytically by a circuit simulator. Finally, in some cases there is no validation of the proposed models with experimental data[39-41]. It is clear that there is still missing a complete approach that combines physics validity with analytical equations that can be easily integrated in a circuit simulator and provide fast and robust solutions.

## Results and Discussion

Thus, the main goal of this work is to propose a physics-based model which accounts for both carrier number and mobility fluctuation noise sources inhomogeneously distributed over the graphene channel and which can be solved analytically. Furthermore, we validate that the developed model can accurately capture the experimentally obtained M-shape gate dependence of 1/f noise data measured in solution-gated GFETs at different bias conditions and for several channel lengths. Residual charge, which is dominant near CNP, will be shown to be responsible for the M-shape dependence, however channel charge inhomogeneity is also found to be significant to the LFN minimum at CNP. As well as this, ΔN is the main 1/f noise contributor for SLG FETs as it was expected[23] but Δμ also contributes near CNP. The contact resistance has a significant effect on 1/f noise at high gate voltages because of the increased and bias dependent contact resistance experimentally observed in this regime[24]. The model also works properly for data from solid-gated GFETs taken from bibliography[30, 32-33]. The basic methodology for the derivation of the physics-based 1/f noise equations in this work is based on a procedure developed for MOSFET devices[15, 19, 43-44]. The implementation of a correct 1/f noise model requires the

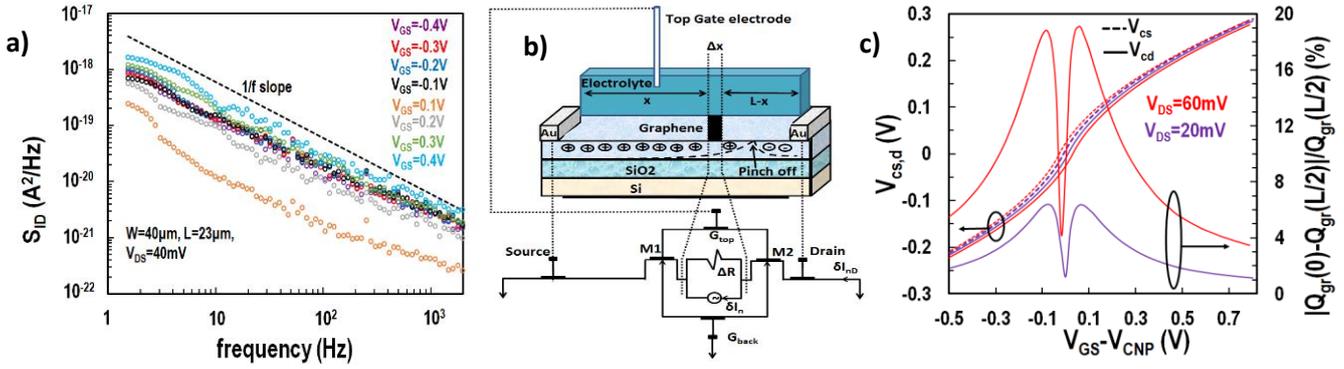

**Fig. 2** a) Relative power spectral density of drain current noise $S_{ID}$ for solution top-gated GFETs with $W=40$ $\mu m$ and $L=23$ $\mu m$ at $V_{DS}=40$ $mV$ for different top gate voltage values ($V_{GS}=-0.4$, $-0.3$, $-0.2$, $-0.1$, $0.2$, $0.3$, $0.4$ $V$); the dashed line corresponds to a 1/f slope. b) Device cross-section. The equivalent circuit for a local current noise contribution to the total noise is illustrated. Each noise-generating slice of the channel is connected to two noiseless GFETs, M1 and M2 respectively. The local current noise source ($\delta I_n$) generates a $\delta I_{nD}$ current fluctuation at the drain. c) Chemical potential $V_c$ (left y-axis) and the relative fluctuation of graphene charge $[Q_{gr}(0) - Q_{gr}(L/2)] / Q_{gr}(L/2)$ (right y-axis) from the beginning ($x=0$) to the middle ($x=L/2$) of the channel are plotted vs top gate voltage overdrive ($V_{GS}-V_{CNP}$) for two drain voltage values of $20$ and $60$ $mV$.

existence of a reliable current–voltage (I-V) model that can qualitatively capture the bias dependence of the drain current of the device. Since LFN expresses the fluctuation of current, thus the absolute current has to be well described. The model for 1/f noise in GFETs has been implemented considering the chemical potential based compact model reported in Refs. 45-46. According to this model, a GFET can be represented by the equivalent capacitive circuit shown in Fig. 1d. Graphene charge $Q_{gr}$ is stored in the quantum capacitance ($C_q$), the chemical potential $V_c(x)$ represents the voltage drop across $C_q$ at position $x$. $V_c(x)$ is defined as the difference between the potential at quasi-Fermi level and the potential at the CNP, as shown in the energy dispersion relation scheme of graphene in Fig. 1d where $V_c(0)=V_{GS}$ at the source end ($x=0$) and $V_c(L)=V_{cd}$ at the drain end ($x=L$). $V_{GS}$-$V_{GSO}$, $V_{BS}$-$V_{BSO}$ are the top and back gate source voltage overdrives while $C_{top}$ and $C_{back}$ are the top and back gate capacitances, respectively. The quasi-Fermi potential $V(x)$ is the voltage drop in the graphene channel at position $x$, which is equal to zero at the source end ($x=0$) and equal to $V_{DS}$ at the drain end ($x=L$).

Drain-to-source current and 1/f noise spectra were measured in single layer, top liquid-gated GFETs with $W=40$ $\mu m$ and five different channel lengths ($L=43$, $23$, $13$, $8$, $5.5$ $\mu m$) (See Experimental Data section). Data were obtained from 4 samples for $L=5.5$, $8$, $23$ $\mu m$, 3 samples for $L=13$ $\mu m$ and 2 samples for $L=43$ $\mu m$, at three different drain voltage levels ($V_{DS}=20$, $40$ and $60$ $mV$). Top gate potential was swept from $V_{GS}=-0.4$ to $0.6V$ with a step of $20$ $mV$, covering the whole range from strong p-type conduction to strong n-type conduction. These extended bias conditions allowed a thorough examination of 1/f noise at all the operation regimes. The measured frequency range from $1.5$ $Hz$ up to $1.5$ $KHz$. Fig. 1e and 1f confirm the excellent agreement of the drain current model and the experiment for all bias and geometry conditions. The compact model reported in Refs 45-46 was used to fit the experimental data obtained from the investigated solution-gated FETs. The values of the model parameters extracted from the fitting of the experimental data are shown in Table 1. The fundamental parameters which are going to be used in noise equations are the carrier mobility ($\mu$), the residual charge density ($\rho_0$), the top gate capacitance ($C_{top}$), the contact resistance

($R_c$) and the flat band top gate voltage ($V_{GSO}$). One parameter set is used for all bias conditions at each channel length; even for different channel lengths, the parameters are quite close to each other. Fig. 2a shows the measured spectra of the $L=23$ $\mu m$ devices at $V_{DS}=40$ $mV$ where it can be observed the 1/f dependence of noise amplitude. LFN can be originated by the local random fluctuations of the carriers' density and the mobility which correspond to the above described $\Delta N$ and $\Delta \mu$ effects, respectively. We develop a physics-based analytical model describing these effects, considering the channel of the device divided into elementary slices[43]. Here, the chemical potential based analytical current model will be used to define the conditions at each channel slice. The fluctuations generating LFN are always small and, consequently, the analysis of the propagation of the noise sources to the voltages or currents at the contact terminals reduces to linear analysis. Therefore, the principle of superposition can be used for adding the effects of the local noise sources along the channel[43]. These local fluctuations can be modeled by adding a random local current noise source $\delta I_n$ with a PSD $S\delta I^2_n$ as shown in Fig. 2b. The local fluctuations propagate to the terminals resulting in fluctuations of the voltages and currents around the DC operating point. The local noise sources are assumed to be spatially uncorrelated and, therefore, their PSDs can be summed. For detailed explanation of the general methodology, see Supplementary Info A. The model considers a non-homogeneous charge distribution along the device channel, according to the physics of GFET, making this approach more realistic. Fig. 2c illustrates, in left y-axis, the chemical potential $V_{cs,d}$ at source and drain terminal respectively, calculated by the employed current model[45-46], vs. top gate voltage overdrive at the lower and higher drain voltage values used in the experiments ($V_{DS}=20$, $60$ $mV$). As predicted by the model, $V_{cd}$ approaches $V_{cs}$ for low $V_{ds}$ values. This effect can be justified from the larger charge homogeneity in the channel at low drain voltage; under these bias conditions $V_c$ is approximately the same at every position in the graphene channel. At $V_{DS}=60$ $mV$, the channel charge non-homogeneity increases with respect to a $V_{DS}=20$ $mV$ and as a result, $V_{cd}$ differs more significantly from $V_{cs}$ especially around CNP (see Fig. 2c).

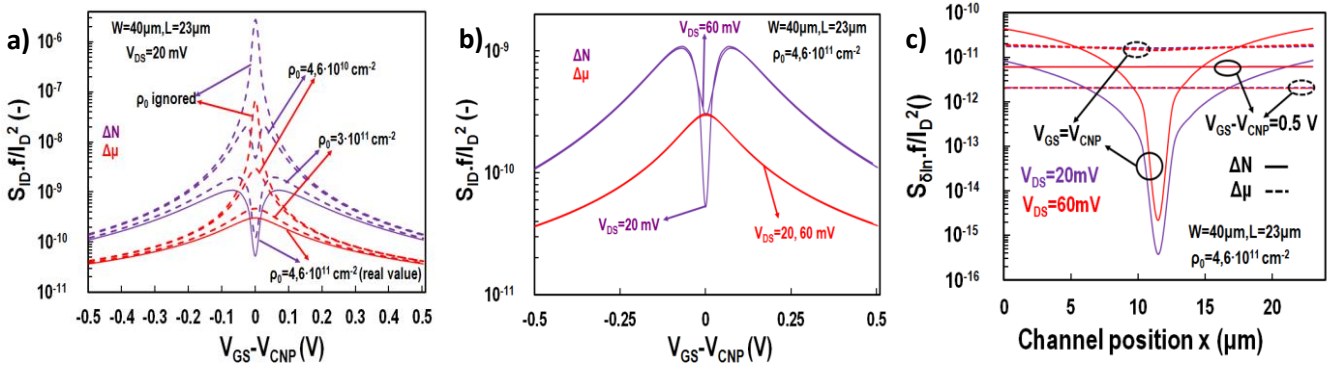

**Fig. 3** Output current noise divided by squared drain current, $S_{ID}/I_D^2$, referred to *1 Hz*, vs. top gate voltage overdrive $V_{GS} - V_{CNP}$ for solution top-gated GFETs. a) $\Delta N$ and $\Delta\mu$ effects at $V_{DS}$ = *20 mV* for four different values of residual charge ($\rho_0$) and for $W/L$=*40 μm/23 μm*. b) $\Delta N$ and $\Delta\mu$ contributions for two $V_{DS}$ values (*20 , 60 mV*). The experimental $\rho_0$ value (*4.6·10⁻¹¹ cm⁻²*) extracted from the current data is used for the calculations. c) Normalized PSD of the local noise, $S_{\delta I_n}/I_D^2$, referred to *1 Hz*, vs. channel position *x*.

At high gate voltages the difference between $V_{cd}$ and $V_{cs}$ becomes less important even for the higher drain voltages, which indicates that the non-homogeneity of the channel is more pronounced near CNP. In the right y-axis of Fig. 2c, the relative fluctuation of $Q_{gr}(x)$ from source terminal to the middle of the channel, $|Q_{gr}(0)-Q_{gr}(L/2)|/Q_{gr}(L/2)$ (%), is shown vs. top gate voltage overdrive for the same drain voltages ($V_{DS}$=*20,60 mV*). The homogeneity of the channel for the small $V_{DS}$ away from $V_{CNP}$ is clear since the observed relative fluctuation of $Q_{gr}(x)$ is insignificant (~1%). As we approach CNP, this fluctuation increases since the channel starts to become non-homogeneous even for this small $V_{DS}$. At $abs(V_{GS}-V_{CNP})$=0.1 V, the relative fluctuation of $Q_{gr}(x)$ reaches its maximum value (~6%) and then it starts to decrease leading to an M-shape behavior similar to that observed in LFN data. This can be justified in terms of the residual charge ($e·\rho_0$=8·10⁻⁸ $C·cm^{-2}$) which starts to contribute to $Q_{gr}(x)$ at this operating point. At $V_{CNP}$, $\rho_0$ is dominant at almost every position of the channel and this leads to the minimum of the relative fluctuation of $Q_{gr}(x)$ observed for the low $V_{DS}$ value (~1%). For the higher $V_{DS}$, an M-shape is also observed for the relative fluctuation of $Q_{gr}(x)$ from source terminal to the middle of the channel but the more intense non-homogeneity leads to higher values. More specifically, the maximum values of the relative fluctuation at $abs(V_{GS}-V_{CNP})$≈0.1 V are almost ~20%. At $V_{CNP}$, the effect of $\rho_0$ decreases the relative fluctuation at a minimum value of ~4% which is significantly higher than the minimum observed at CNP for the lower $V_{DS}$. This also occurs because of the inhomogeneity of the channel at the higher $V_{DS}$. For detailed explanation of the behaviour and value of $Q_{gr}(x)$ at every channel position *x* under different bias conditions, see Supplementary Info B (Figure S1).

Considering the carrier number fluctuation effect, if a certain number of carriers is trapped at channel position *x*, the relative current fluctuation can be calculated as:

$$\frac{\delta I_D(x)}{I_D} = \frac{\delta N_{gr}}{N_{gr}} = \frac{1}{N_{gr}} \cdot \frac{\partial Q_{gr}}{\partial Q_t} \cdot \partial N_t \qquad (1)$$

where $N_{gr}$ is the graphene carrier density and $Q_t$, $N_t$ are the trapped charge and density respectively; charges and number of carriers are expressed per unit area since they are referred to a channel slice. Fluctuations of the trapped charge $\delta Q_t$ can cause a variation in the chemical potential $\delta V_c$ which can lead to a change of charges that depend directly on the chemical potential such as the graphene charge, the top gate and the back gate charge. By applying charge conservation law and by considering a linear dependence of the quantum capacitance $C_q$ and the chemical potential $V_c$ ($C_q$=$k·|V_c|$)[45-46], with k defined in Supplementary Info A, the following expression is derived:

$$\frac{\delta I_D(x)}{I_D} = \frac{e}{Q_{gr}} \cdot \frac{C_q}{C_q + C_{top} + C_{back}} \cdot \partial N_t \qquad (2)$$

and the PSD of the local noise source normalized by squared drain current can then be calculated as (see Supplementary Info A):

$$\frac{S_{\delta I_n^2}}{I_D^2}\Big|_{\Delta N} = \left(\frac{e}{Q_{gr}} \cdot \frac{C_q}{C_q + C_{top} + C_{back}}\right)^2 \frac{KT\lambda N_T}{W\Delta xf} \qquad (3)$$

, where $N_T$ is the dielectric volumetric trap density per unit energy (in $eV^{-1}cm^{-3}$) which is used as a fitting parameter, *K* is the Boltzmann constant, *T* is the absolute temperature, *e* the electron charge, $\lambda$~0.1 nm is the tunneling attenuation distance since the trapping/detrapping mechanism is considered a tunneling process. The analysis of this process is difficult at atom level, thus the best possible approach is to model the capture cross-section according to $P(z) = \exp(-z/\lambda)$, where *P* is the tunnelling probability of a carrier to get captured by a trap located at a barrier depth *z* into the dielectric. $C_{back}$ is not defined for the measured devices in this work but is included in the equations for better generalizability of the proposed model. By integrating the PSD of the local noise source in the entire channel length[41] and by changing the integration variable from length to chemical potential at source and drain terminals[45-46], it is possible to derive the following analytical formula for the relative PSD of the total fluctuation of the drain current resulting from a carrier density fluctuation $\Delta N$:

$$\frac{S_{I_D}}{I_D^2}f\Big|_{\Delta N} = \frac{SD|_{\Delta N} \cdot KD|_{\Delta N}}{\left[g(V_c)\right]_{V_{cs}}^{V_{cd}}} \qquad (4)$$

$SD|_{\Delta N}$=$2·KT·\lambda·N_T·e^2/(C·WL·k)$ is a bias independent term representing the amplitude of the $\Delta N$ effect noise, where $C$=$C_{top}+C_{back}$. $KD|_{\Delta N}$ is a bias dependent term of the $\Delta N$ model and

$$\text{for } V_{cs,cd} > 0, \quad KD\Big|_{\Delta N} = \frac{1}{(\alpha k + C^2)}\left[\alpha k \ln\left(\alpha + kV_c^2\right) - 2\alpha k \ln(\alpha) + 2C^2\ln\left(C + kV_c\right) - 4C^2\ln(C) - 2\sqrt{\alpha k}C\cdot\arctan\left(\sqrt{\frac{k}{a}}V_c\right)\right]_{V_{cd}}^{V_{cs}}$$

$$\text{for } V_{cs,cd} < 0, \quad KD\Big|_{\Delta N} = \frac{1}{(\alpha k + C^2)}\left[-\alpha k \ln\left(\alpha + kV_c^2\right) - 2C^2\ln\left(C - kV_c\right) - 2\sqrt{\alpha k}C\cdot\arctan\left(\sqrt{\frac{k}{a}}V_c\right)\right]_{V_{cd}}^{V_{cs}}$$

(5)

is defined by Equation (5), where $\alpha = 2 \cdot \rho_0 \cdot e$. Finally, $g(V_c)$ is a bias dependent term proportional to the drain current[45-46] (see also Supplementary Info A). As far as the mobility fluctuation effect is concerned, by using a methodology identical to the presented above, the following analytical formula is obtained:

$$\frac{S_{I_D}}{I_D^2}f\Big|_{\Delta\mu} = \frac{SD\big|_{\Delta\mu}\cdot KD\big|_{\Delta\mu}}{\left[g(V_c)\right]_{V_{cs}}^{V_{cd}}}$$

(6)

$$\text{for } V_{cs,cd} > 0, \quad KD\Big|_{\Delta\mu} = \left[CV_c + \frac{kV_c^2}{2}\right]_{V_{cd}}^{V_{cs}}$$

$$\text{for } V_{cs,cd} < 0, \quad KD\Big|_{\Delta\mu} = \left[CV_c - \frac{kV_c^2}{2}\right]_{V_{cd}}^{V_{cs}}$$

(7)

(see Supplementary Info A). The bias dependent term $KD/_{\Delta\mu}$ is given by Equation (7) where residual charge related term $\alpha$ does not play any role and $SD/_{\Delta\mu} = 2\cdot\alpha_H\cdot e/(C\cdot WL\cdot k)$ where $\alpha_H$ is the unitless Hooge parameter which is used as a fitting parameter. In order to calculate the total $1/f$ noise of the device, the two different contributions have to be added as:

$$\frac{S_{I_D}}{I_D^2} = \frac{S_{I_D}}{I_D^2}\Big|_{\Delta N} + \frac{S_{I_D}}{I_D^2}\Big|_{\Delta\mu}$$

(8)

The strong dependence of $1/f$ noise on both residual charge and channel charge inhomogeneity makes it essential to thoroughly investigate these phenomena. Fig.3a illustrates the dependence of the two $1/f$ noise models, $\Delta N$ and $\Delta\mu$, on the residual charge and Fig. 3b on the drain voltage. In Fig. 3a the contributions of both noise mechanisms $\Delta N$ and $\Delta\mu$ are shown for different values of the residual charge density $\rho_0$ at $V_{DS}$=20 mV. The value ($4,6\cdot10^{11}$ cm$^{-2}$) corresponds to the value experimentally extracted from fitting the I-V data (see Fig. 1e and 1f). In addition, the model is tested at three other lower values of $\rho_0$ ($3\cdot10^{11}$, $4,6\cdot10^{10}$, o cm$^{-2}$). It can be concluded from Fig. 3a that the $\Delta N$ effect is responsible for the M-shape bias dependence in case of relatively high $\rho_0$ values are considered (see Fig. 2c) while for low $\rho_0$ values, a $\Lambda$ shape behavior is obtained. $\Delta\mu$ model always provides an $\Lambda$ shape behavior with an increased maximum at CNP as $\rho_0$ decreases since $\rho_0$ only affects normalized drain current term $g(V_c)$ in

Equation 6 and not the bias dependent term $KD/_{\Delta\mu}$ of Equation (7). (For more information, see Supplementary Info C). Regarding the drain voltage dependence, Fig. 3b indicates that the increase of $V_{DS}$ (20, 60 mV) increases the contribution of the $\Delta N$ noise near the CNP resulting from the increased graphene charge inhomogeneity observed at higher $V_{DS}$ (see Fig. 2c) while any $\Delta\mu$ noise remains unaffected. At higher gate voltages no drain voltage dependence can be observed by any of the noise mechanisms confirming that $Q_{gr}$ is homogeneous at high $V_{GS}$ values. According to Fig. 2c, $V_{cs}$, $V_{cd}$ are very close for higher $V_{GS}$ values and since Equations 4-7 show that the bias dependence of both noise mechanisms $\Delta N$ and $\Delta\mu$, is exclusively expressed in terms of chemical potentials $V_{cs}$, $V_{cd}$, the drain voltage independence of both $\Delta N$ and $\Delta\mu$ noise mechanisms for higher gate voltages can be explained. In Fig. 3c, the local noise at each channel position $x$ is shown at $V_{GS}=V_{CNP}$ and at $V_{GS}-V_{CNP}$=0,5 V for both noise mechanisms $\Delta N$ and $\Delta\mu$, as it is calculated by Equations (3) and (A12) respectively for $V_{DS}$=20, 60 mV. At $V_{CNP}$, the total noise $\Delta N$ propagated to the terminals is mainly determined by the local noise at the source/drain ends while away from CNP, all the points along the channel contributes equally. This proves the homogeneity of the channel at higher gate voltages while the different contributions of the charge distributed along the channel at $V_{CNP}$ indicate the channel inhomogeneity close to CNP, especially for the higher $V_{DS}$, as described in Fig. 2c. Regarding $\Delta\mu$ noise, all the points of the channel contribute similarly at every bias condition. By summing the local $\Delta N$ and $\Delta\mu$ noise sources throughout the channel, we can accurately obtain the values of the total $\Delta N$ and $\Delta\mu$ noise PSD as calculated by Equations (4) and (6) and as shown in Fig. 3b for the operating conditions under study. The effect of $\rho_0$ in the local LFN is shown in Supplementary Info C.

Fig. 4 shows the experimental noise data averaged in the bandwidth of $10 - 40$ Hz, referred to $1$ Hz. The data are fitted using the same parameters extracted from the current compact model and adjusting only the $N_T$ and $\alpha_H$ values. Fig. 4a and 4b present the normalized noise data for transistors with two different channel lengths, $L$=43 μm and $L$=5.5 μm, respectively, at two drain voltage values (20 and 60 mV). Fig. 4c shows the fitted normalized noise data for two other channel lengths (23 and 8 μm) at all the drain voltage values (20, 40, 60 mV) (see Fig. S3 in Supplementary Info D for the complete set of data). The symbols correspond to the experimental data and the solid lines represent

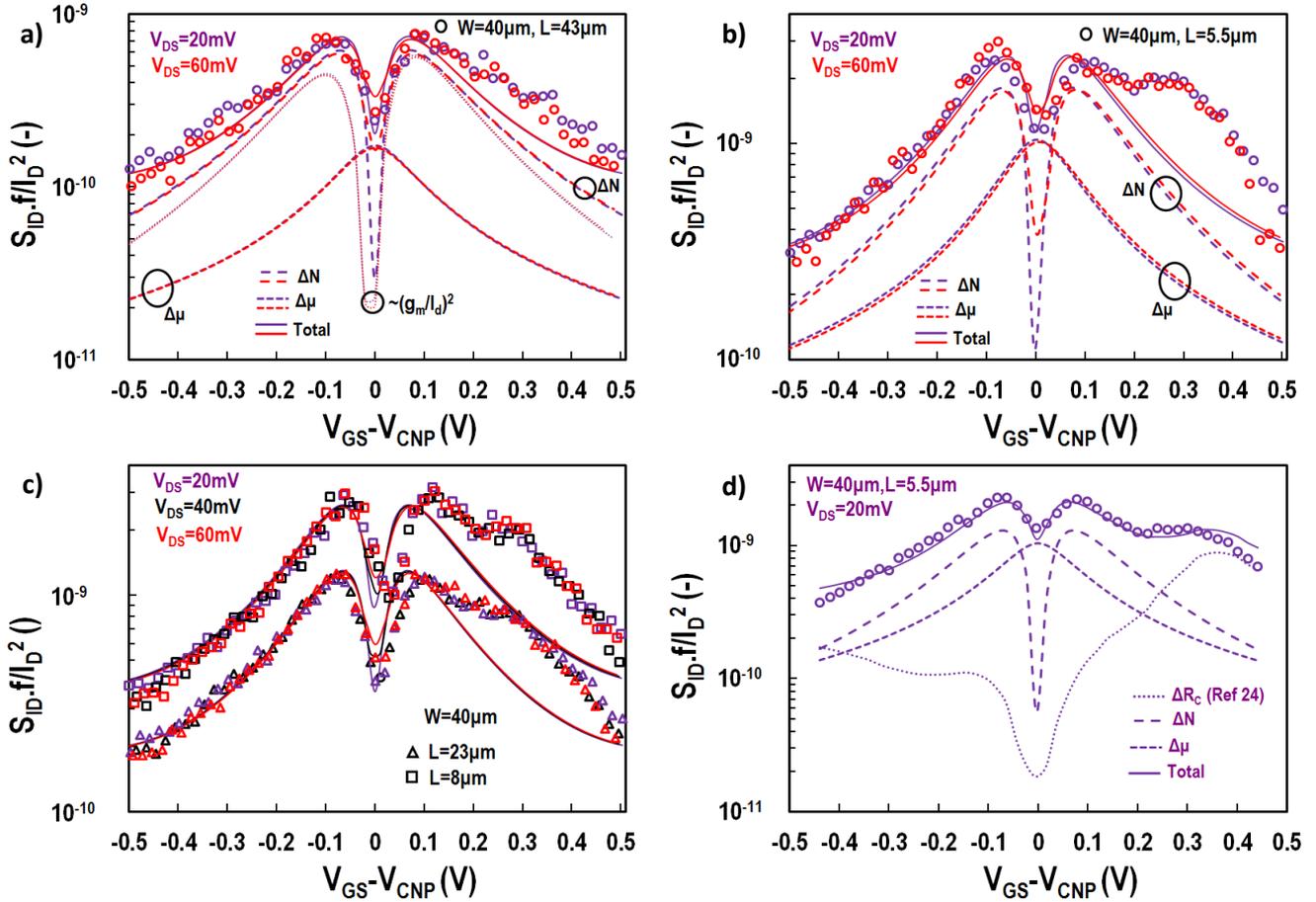

**Fig. 4** Output noise divided by squared drain current $S_{ID}/I_D{}^2$, referred to *1 Hz*, vs. top gate voltage overdrive ($V_{GS} − V_{CNP}$), for solution top-gated GFETs with *W=40 μm*. a) Data from transistors with a channel length *L=43 μm* and *V_{DS} = 20* and *60 mV*. The experimental data is fitted with the derived compact model (the *ΔN* and *Δμ* contributions are displayed separately). The simplified $(g_m/I_D)^2$ noise model[4, 16-17] (Supplementary Info E, Equation A24), which considers charge homogeneous along the channel is plotted with dotted lines for both *V_{DS}* values. b) Data from transistors with an *L=5.5 μm* and *V_{DS} = 20* and *60 mV*. A fitting of the data and the individual contributions from *ΔN* and *Δμ* are also plotted. c) Data from transistors with a channel length *L=23 μm and 8 μm* at *V_{DS} = 20, 40* and *60 mV* is plotted together with the fitting provided by the analytical model. d) Experimental data from transistors with *L=5.5 μm* and *V_{DS} = 20mV* is shown. A contact noise term[23] (dotted line) is added to the compact model to correct the deviations from experimental data away from the CNP.

the total 1/f noise model. The well-known M-shape trend is observed in our data near the CNP. The change in the minimum value at the CNP with *V_{DS}* caused by the charge inhomogeneity is also properly described. Away from the CNP, the measured noise is independent on the drain voltage and the model follows this trend as well. Dashed lines representing the different 1/f noise contributors in Fig. 4a and 4b, provide additional insights on the contributions of the different noise mechanisms. The dotted lines in Figure 4a present the simplified *(g_m/I_D)²* model[16-17] (See Supplementary Info E, Figure S4) for both drain voltages available. It is apparent that the specific approach cannot capture the drain voltage dependence of LFN near CNP since it considers a uniform charge along the channel. The *ΔN* mechanism is responsible for the M-shape, as it was shown previously in Fig. 3a. Despite the fact that the *ΔN* model can predict the drain voltage dependence near CNP, it significantly underestimates the minimum of noise near the CNP. On the other hand, the *Δμ* model predicts a *Λ*-shape dependence with the gate bias which is independent on the drain voltage. This term can have a significant effect near CNP, setting a minimum noise value that helps to fit better the experimental data (see Fig. 4a and 4b). The distinction of the *ΔN* and *Δμ* contributions near the CNP is shown in this work for the first time. The normalized noise increases with decreasing device area, as it is apparent in Figure 4c; this is expected since 1/f noise is known to scale inversely proportional with the device dimensions. As it can be derived from Equations (4) and (6). The higher noise measured in the n-type conduction regime, more pronounced at higher gate voltages and at shorter channel lengths, is tentatively attributed to the bias dependent contact resistance experimentally observed in this bias regime[24]. Fig. 4d shows the model corrected to include a contact noise contribution[24] as reported previously. To calculate the magnitude of this contribution, the contact resistance has been calculated using a transmission line method (TLM) analysis. The contact noise model used to refine the fitting of the experimental noise also proves that contact noise is negligible near the CNP. All the extracted 1/f noise parameters are shown in Table 1; it is important to highlight that for a fixed channel length, the same parameters are used to fit the whole range of bias conditions. Regarding the level of normalized 1/f noise, the values of the extracted noise

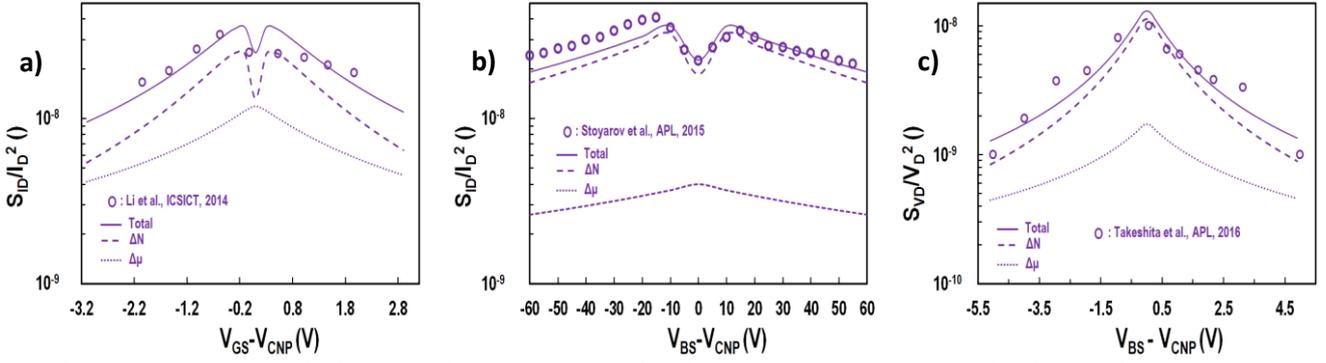

**Fig. 5** a): Output current noise divided by squared drain current $S_{ID}/I_D^2$, referred to _1 Hz_, vs. top gate voltage overdrive ($V_{GS} - V_{CNP}$), for a top-gated GFETs with $W/L=12\ \mu m/0.35\ \mu m$ at $V_{DS} = 0.2\ V$ (Ref. 30). b): Output current noise divided by squared drain current $S_{ID}/I_D^2$, referred to _1 Hz_, vs. back gate voltage overdrive ($V_{BS} - V_{CNP}$), for a back-gated GFETs with $W/L=6.3\ \mu m/2.1\ \mu m$ at $V_{DS} = 0.01\ V$ (Ref. 32). c) Output voltage noise divided by squared voltage potential $S_{VD}/V_D^2$, referred to _1 Hz_, vs. back gate voltage overdrive ($V_{BS} - V_{CNP}$), for a back-gated GFETs with $W/L=12.7\ \mu m/11\ \mu m$ at $V_{DS} \sim 0.6\ mV$ (Ref. 33). Symbols: data[30, 32-33], solid line: model, dashed lines: different noise contributions, $\Delta N$ and $\Delta\mu$.

**Table 1. Drain Current and 1/f Noise model Parameters**

| Parameter | Units | L=43µm | L=23µm | L=13µm | L=8µm | L=5.5µm | Ref. 30 | Ref. 32 | Ref. 33 |
|---|---|---|---|---|---|---|---|---|---|
| $\mu$ | cm²/(V·s) | 3400 | 3250 | 3400 | 3400 | 3600 | 950 | 3000 | 4500 |
| $C_{top}$ | µF/cm² | 1,9 | 1,9 | 2,05 | 2,05 | 2,2 | 0,65 | $C_{back}$=1,9·10⁻² | $C_{back}$=1,15.10⁻² |
| $V_{GSO}$ | V | 0,09 | 0,09 | 0,095 | 0,095 | 0,095 | 0 | $V_{BSO}$=7,75 | $V_{BSO}$=4 |
| $\rho_0$ | cm⁻² | 5·10¹¹ | 4,6·10¹¹ | 5,3·10¹¹ | 5·10¹¹ | 4,8·10¹¹ | 5,9·10¹² | 1,3·10¹³ | 1,2·10¹¹ |
| $R_c$ | Ω | 120 | 120 | 120 | 120 | 120 | - | - | - |
| $N_T$ | eV⁻¹cm⁻³ | 1,3·10²⁰ | 1,1·10²⁰ | 1·10²⁰ | 9·10¹⁹ | 5,5·10¹⁹ | 3·10²⁰ | 3,5·10²¹ | 2·10²⁰ |
| $\alpha_H$ | - | 1,5·10⁻³ | 1,3·10⁻³ | 1,3·10⁻³ | 1,2·10⁻³ | 1,1·10⁻³ | 3·10⁻³ | 7·10⁻³ | 3·10⁻⁴ |

parameters are in the same order of magnitude or lower than what is available in bibliography for graphene devices. The $\alpha_H$ parameter is lower than many reports[28, 30, 37] even considering that the Hooge model underestimates 1/f noise, since the $\Delta N$ effect is more dominant as shown in Fig. 4. In some reports[4, 42], the $N_T$ parameter is also derived and it is quite comparable with the values in Table 1; still, $N_T$ is higher than its typical range at Si devices ($N_T\sim10^{16}$-$10^{18}\ eV^{-1}cm^{-3}$)[19, 43]. The noise amplitude $B=f\cdot Area\cdot S_{ID}/I_D^2$, can be easily found to range from $10^{-7}\sim10^{-6}\ \mu m^2$ in the present work (See Figure S5 in Supplementary Info F), which is similar or lower in comparison with other works[4, 28-29, 33, 37-38].

In order to confirm the generalizability of the proposed model, we have tested it with datasets of three solid gated GFETs taken from literature[30, 32-33]. Fig. 5a shows the $S_{ID}/I_D^2$ noise data for a top-gated device with $W/L=12\ \mu m/0.35\ \mu m$ from Ref. 30 (Fig. 4b[30], _T=300 K_) at $V_{DS}=0.2\ V$, Fig. 5b presents the $S_{ID}/I_D^2$ noise data for a back-gated device with $W/L=6.3\ \mu m/2.1\ \mu m$ from Ref. 32 (Fig. 4a[32], _Si/SiO2_ data) at $V_{DS}=0.01\ V$ and Fig. 5c illustrates the $S_{VD}/V_D^2$ 1/f noise data for a back-gated GFET with $W/L=12.7\ \mu m/11\ \mu m$ from Ref. 33 (Fig. 3b[33], _T=1.6 K_) at $V_{DS}\sim0.6\ mV$. The two different representations of normalized noise displayed in Fig. 5 ($S_{ID}/I_D^2$ and $S_{VD}/V_D^2$) are equivalent. The symbols represent the measurements and the total model is shown by the solid lines, the $\Delta N$ and $\Delta\mu$ contributions are also shown with dashed and dotted lines respectively. Regarding Fig. 5a where the M-shape dependence of noise is also observed, the total model behavior is acceptable. Additionally, both $\Delta N$ and $\Delta\mu$ effects have a significant contribution especially near CNP, similarly as in Fig. 4. In Fig. 5b, the M-shape dependence of noise is intense probably due to a higher residual charge value and our model captures well this shape. $\Delta N$ effect is the dominant noise source while $\Delta\mu$ effect has a small contribution near CNP. The LFN data in Fig. 5b are asymmetrical with an increased value at p-type region while our model is equivalent in both n- and p- regions. We extracted the noise parameters by targeting a better performance in n-type conduction but we could achieve an overall better fitting by using different LFN parameters below and above CNP. Finally, Fig. 5c shows that the 1/f normalized noise follows a $\Lambda-shape$ behavior. By fitting the noise curve with our model, it is possible to distinguish between the $\Delta N$ and the $\Delta\mu$ effects due to the different slopes of their curves. This finding can be explained by the relatively small value of the residual charge $\rho_0$=1,2·10¹¹ cm⁻³ in this device. The parameters extracted are also presented in the last three columns of Table 1. For all devices, the $N_T$ parameter is a little higher than the ones extracted for our dataset. Regarding the $\alpha_H$ parameter, it is in the same level as in our data set for the plot in Fig. 5a while it is quite lower in the Fig. 5b but in this case the error of the fitting can be quite significant.

## Conclusions

In conclusion, this paper investigates the bias-dependence of 1/f noise in liquid gated, single layer GFETs. An analytical compact model is developed considering both carrier number and mobility fluctuation mechanisms. According to this procedure, the noise in an elementary slice of the channel is calculated based on physical laws; after integrating along the channel, the local noise sources are propagated to the terminals and the final formulas are derived. In this compact format the model can be easily implemented in Verilog-A code and integrated in circuit simulators, which could be instrumental to bridge the gap between device and circuit levels. The model is capable of quantitatively capture the experimental M-shape of normalized output noise which is observed for all channel lengths and drain voltages available. The simultaneous contribution of the $\Delta N$ and $\Delta \mu$ noise mechanisms significantly improves the prediction accuracy of the model, confirming that both noise contributions are needed to properly model noise in graphene FETs. Additionally, a previously reported contact noise term based on carrier number fluctuations proved to be effective to account for such contribution. An analytical solution of the LFN generated by contact resistance is an essential future step so as our model to be capable of capturing additional behaviours of LFN mentioned in bibliography such as an extended V-shape vs gate voltage and thus, to be considered complete. The analytical model presented in this work encompasses all the main contributions to 1/f noise in graphene FETs, taking into account the non-homogeneities in the channel. Such an analytical and yet complete model can be of high interest to identify and understand the main causes of noise as well as for boosting the design of integrated circuits based on graphene.

## Experimental Data

### Graphene CVD growth and transfer

Graphene is synthesized by chemical vapour deposition (CVD) technique on a copper foil. A chemical wet transfer method is used to transfer the graphene from the Cu foil to the $SiO_2$ substrate. First, the graphene is protected with a sacrificial poly (methyl meth-acrylate) PMMA layer. Subsequently, the back side graphene is removed by oxygen plasma treatment. The Cu foil is then placed in $FeCl_3$ 0.5 M/ HCl 2 M (1:2) etchant solution until all the Cu is chemically dissolved. Before the final transfer onto the desired $SiO_2$ substrate, the graphene/PMMA stack is placed several times in DI water to rinse the residual etchant solution away. The wafer is dried for 30 minutes at 40 ºC on a hot plate and then gradually increased up to 180 ºC in a vacuum oven. Finally, the PMMA is removed in acetone and IPA.

### Devices fabrication

Arrays of graphene transistors are fabricated on 4-inch $Si$/ $SiO_2$ wafer with a 285 nm thick layer of thermal silicon oxide. A first metal layer of Ti/Au is deposited by electron-beam evaporation and structured by a lift-off process. Afterwards, the CVD-gown graphene is transferred as previously described. The graphene transistor active area is protected by a photo definable resist HIPR 6512. Thus, graphene is patterned by oxygen plasma in a reactive ion etching (RIE) system. Top contacts of Ni/Au are deposited by evaporation and defined by lift-off. In order to prevent any damage of graphene, the lift-off is performed by leaving the wafer

1 hour in acetone and flushing it with a syringe. After 2 hours annealing step at 300 ºC in ultra-high vacuum, a 2 μm thick SU8 negative epoxy resist (SU-8 2005 MicroChem) layer is spin coated and structured such that only the graphene between source and drain contact is exposed to the electrolyte.

### Electrical characterization

The liquid-gated graphene-transistor characteristics are measured in a 10 mM PBS electrolyte. The gate voltage is applied versus an Ag/AgCl reference electrode. At each polarization, the drain-to-source current signal is measured with a custom-made current-to-voltage converter with two parallel inputs for DC (low-pass filter at 0. 1Hz for I-V characteristics) and AC (band-pass filter from 0.1 Hz to 7 kHz for noise characterization). The data acquisition is performed using a National Instruments DAQ-card system (NI 6363). In order to stabilize the $I_{DS}$ current value at each gate bias, the sampling condition is $dI_{DS}/dt < 1 \cdot 10^7$ A/s before each recorded point. For the noise characterization, the sampling frequency was set to 50 kHz for a period of time of 13 seconds choosing the Welch's method in which 10 segments overlap by 50%.

### Data availability

The data that support the findings of this study are available from Ramon Garcia Cortadella and Andrea Bonaccini Calia. Please, address your requests to ramon.garcia@icn2.cat and andrea.bonaccini@icn2.cat.

## Acknowledgements

This work was funded by the Ministerio de Economía y Competitividad under the project TEC2015-67462-C2-1-R and the European Union's Horizon 2020 research and innovation program under Grant Agreement No. GrapheneCore2 785219 (Graphene Flagship), Marie Skłodowska-Curie Grant Agreement No 665919 and Grant Agreement No. 732032 (BrainCom).

**Supplementary Information for:**

# Understanding the Bias Dependence of Low Frequency Noise in Single Layer Graphene FETs

Nikolaos Mavredakis*[a], Ramon Garcia Cortadella[b], Andrea Bonaccini Calia[b], Jose A. Garrido[b] and David Jiménez[a]

[a] Departament d'Enginyeria Electrònica, Escola d'Enginyeria, Universitat Autònoma de Barcelona, Bellaterra 08193, Spain

[b] Catalan Institute of Nanoscience and Nanotechnology (ICN2), CSIC, Barcelona Institute of Science and Technology, Campus UAB, Bellaterra, Barcelona, Spain

* nikolaos.mavredakis@uab.es

### A. Supplementary Information: Thorough theoretical procedure for equations extraction

### Generalized Noise Modeling methodology:

Under the assumption that the channel of the device is noiseless apart from an elementary slice between positions $\chi$ and $\chi+\Delta\chi$ as it is shown in Fig. 2b in the manuscript, the microscopic noise coming from this slice of the channel can be modeled as a local current source $\delta I_n$ with a PSD $S_{\delta I^2_n}$ which is connected between $\chi$ and $\chi+\Delta\chi$ in parallel with the resistance of the slice $\Delta R$ (Norton equivalent) [43]. The transistor then can be split into two noiseless transistors M1 and M2 on each side of the local current noise source, at the source and drain side ends with channel lengths equal to $\chi$ and $L$-$\chi$ respectively. Since the voltage fluctuations on parallel resistance $\Delta R$ are small enough compared to thermal voltage $U_T$, small signal analysis can be used in order to extract a noise model according to which, M1 and M2 can be replaced by two simple conductances $G_S$ on the source side and $G_D$ on the drain side. The total channel conductance comes from the series connection of $G_S$ and $G_D$ as: $1/G_{CH}=1/G_S+1/G_D$ [43]. The fluctuation of the current due to the local current noise source at the drain side $\delta I_{nD}$ and its corresponding PSD $S_{\delta I^2_{nD}}$ are given by the following equations [43]:

$$\delta I_{nD} = G_{CH}\Delta R\delta I_n \qquad \text{(Eq. A1)}$$

$$S_{\delta I^2_{nD}}(\omega, x) = G^2_{CH}\Delta R^2 S_{\delta I^2_n}(\omega, x) \qquad \text{(Eq. A2)}$$

The PSD of the total noise current fluctuation at the drain side $S_{ID}$ due to all different sections along the channel is obtained by summing their elementary contributions $S_{\delta I^2_{nD}}$ assuming that the contribution of each slice at different positions along the channel remains uncorrelated [43]:

$$S_{ID} = \int_0^L G_{CH}^2 \Delta R^2 \frac{S_{\delta I_2^2}(\omega, x)}{\Delta x} dx = \frac{1}{L^2} \int_0^L \Delta x S_{\delta I_2^2}(\omega, x) dx, \quad where \ G_{CH}^2 \Delta R^2 = \left(\frac{\Delta x}{L}\right)^2 \qquad \text{(Eq. A3)}$$

**Carrier Number Fluctuation Effect:**

As mentioned in the manuscript, the fluctuation of the trapped charge $\delta Q_t$ can cause a variation in the chemical potential $\delta V_c$ which can lead to a change to all charges that depend directly on chemical potential such as the graphene charge, the top gate and the back gate charge. The application of the charge conservation law gives:

$$\delta Q_{gr} + \delta Q_{top} + \delta Q_{back} + \delta Q_t = 0 \qquad \text{(Eq. A4)}$$

These induced fluctuations of the graphene, top gate and back gate charges can be related to the fluctuation of the chemical potential $\delta V_c$ as[15, 43-46]:

$$\delta Q_{gr} = -C_q \delta V_c$$
$$\delta Q_{top} = -C_{top} \delta V_c \qquad \text{(Eq. A5)}$$
$$\delta Q_{back} = -C_{back} \delta V_c$$

If eqns (A4, A5) are taken into account then eqn (1) is transformed in eqn (3) in the manuscript. If the linear relationship between quantum capacitance and chemical potential mentioned in the manuscript, is integrated, charge of graphene can be calculated as:

$$Q_{gr} = \frac{k \cdot V_c^2}{2} + \rho_0 \cdot e \qquad \text{(Eq. A6)}$$

The PSD of the local noise source is calculated by eqn (4) in the manuscript. Taking the integral of this from Source to Drain in order to calculate the total 1/f noise PSD as in eqn (A3)[15, 43], we have:

$$\frac{S_{I_D}}{I_D^2} f \bigg|_{\Delta N} = \frac{1}{L^2} \int_0^L \left(\frac{e}{Q_{gr}}\right)^2 \left(\frac{C_q}{C_{top} + C_{back} + C_q}\right)^2 \cdot \frac{KT\lambda N_T}{W} dx \qquad \text{(Eq. A7)}$$

In order to express this integral in terms of chemical potential $V_c$, we have to change the integral variable as[45-46]:

$$\frac{dx}{dV_c} = \frac{-\mu W Q_{gr}}{I_D} \frac{C_q + C_{top} + C_{back}}{C_{top} + C_{back}} \qquad \text{(Eq.A8)}$$

Where drain current is given as[45-46]:

$$I_D = \frac{\mu W k}{2L} \left[ g\left(V_C\right) \right]_{V_{cs}}^{V_{cd}} \qquad \text{(Eq.A9)}$$

With $k=2\cdot e^3/(\pi\cdot h^2\cdot v^2 f)$[45-46] where $vf$ is the Fermi velocity ($=10^6$ m/s) and $h$ the reduced Planck constant ($=1,05\cdot10^{-34}$ J·s). Bias dependent term $g(V_c)$ is calculated as[45-46]:

$$\left[ g\left(V_C\right) \right]_{V_{cs}}^{V_{cd}} = \frac{V_{cs}^3 - V_{cd}^3}{3} + \frac{k}{4\left(C_{top} + C_{back}\right)}\left[ \text{sgn}\left(V_{cd}\right)V_{cd}^4 - \text{sgn}\left(V_{cs}\right)V_{cs}^4 \right] + \frac{2\rho_0 e V_{DS}}{k} \qquad \text{(Eq.A10)}$$

eqn (A7) is transformed because of eqns (A8, A9, A10) to:

$$\frac{S_{I_D}}{I_D^2} f \bigg|_{\Delta N} = \frac{4KT\lambda N_T e^2 k}{WL\left[ g\left(V_C\right) \right]_{V_{cs}}^{V_{cd}}\left(C_{top} + C_{back}\right)} \int_{V_{cd}}^{V_{cs}} \frac{V_c^2}{\left(kV_c^2 + 2\rho_0 e\right)\left(C_{top} + C_{back} + k|V_c|\right)} dV_c \qquad \text{(Eq.A11)}$$

The integral in eqn (A11) can be solved analytically and gives the eqns (2, 5) in the manuscript.

**Mobility Fluctuation Effect:**

In the empirical Hooge model, the PSD of the local noise source is expressed as[43]:

$$\frac{S_{\delta i_n^2}}{I_D^2} \bigg|_{\Delta\mu} = \frac{\alpha_H e}{Q_{gr}W\Delta x f} \qquad \text{(Eq.A12)}$$

If eqn (A12) is integrated along the channel as eqn (A3), the total noise PSD due to mobility fluctuations effect can be calculated as[43]:

$$\frac{S_{I_D}}{I_D^2} f \bigg|_{\Delta\mu} = \frac{\alpha_H e}{WL^2} \int_0^L \frac{1}{Q_{gr}} dx \qquad \text{(Eq. A13)}$$

If eqn (A8) is applied in order to change the integration variable from $x$ to $V_c$:

$$\frac{S_{I_D}}{I_D^2} f \bigg|_{\Delta\mu} = \frac{\alpha_H e}{L^2 W\left(C_{top} + C_{back}\right)} \int_{V_{cd}}^{V_{cs}} \frac{\mu W Q_{gr}}{Q_{gr} I_D}\left(k|V_c| + C_{top} + C_{back}\right) dV_c \qquad \text{(Eq.A14)}$$

Where $Q_{gr}$ is simplified in eqn (A14) and does not play a role in mobility fluctuation effect. If eqns (A9, A10) are also taken into account, then:

$$\frac{S_{I_D}}{I_D^2} f \bigg|_{\Delta\mu} = \frac{2\alpha_H e}{WLk\left[ g\left(V_C\right) \right]_{V_{cs}}^{V_{cd}}\left(C_{top} + C_{back}\right)} \int_{V_{cd}}^{V_{cs}} \left(k|V_c| + C_{top} + C_{back}\right) dV_c \qquad \text{(Eq.A15)}$$

The integral in eqn (A12) can be solved analytically and gives the eqns (6, 7) in the manuscript.

**B. Supplementary Information Figure S1: Detailed examination of graphene charge along the channel.**

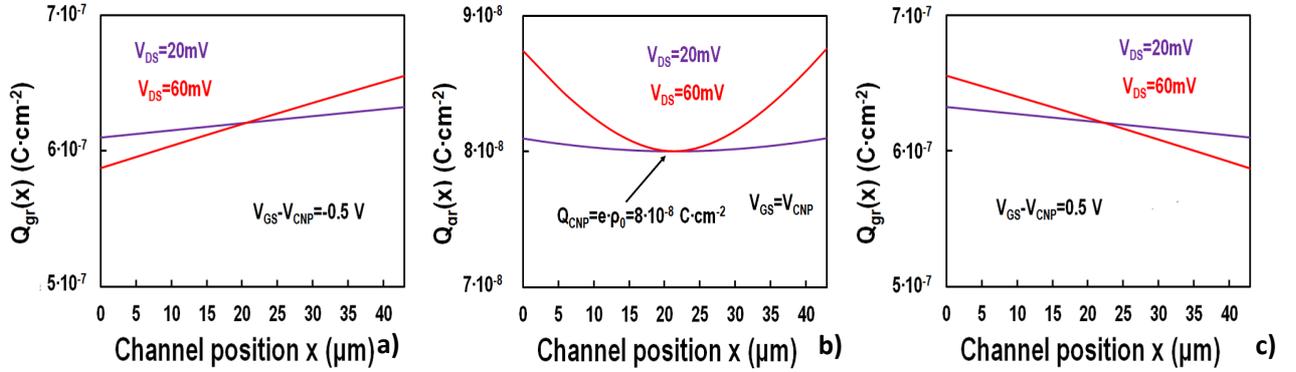

Figure S1. Graphene charge $Q_{gr}(x)$ vs. channel position $x$, for $V_{GS}-V_{CNP}$ = -0,5 V (a), 0 V (CVP) b) and 0,5 V (c) at $V_{DS}$ = 20, 60 mV for W/L=40 μm/43 μm.

Away from CNP (Fig. S1a-S1c), $Q_{gr}(x)$ is ~6-6.5·10$^{-7}$ C.cm$^{-2}$ all along the channel for both drain voltage values. Considering the relative fluctuation of $Q_{gr}(x)$ from source terminal to the middle of the channel shown in Fig. 2c of the manuscript, the homogeneity of the channel is shown at high gate voltages for both $V_{DS}$ values. Near CNP (Fig. S1b), $Q_{gr}(x)$ is equal to residual charge, $e·\rho_0$, at $x=L/2$ for both high and low $V_{DS}$. This value remains almost constant throughout the channel for $V_{DS}$=20 mV but it is increased significantly for $V_{DS}$=60 mV confirming the inhomogeneous channel under these conditions.

**C. Supplementary Information: Detailed examination of effect of residual charge in the M-shape bias dependence of 1/f noise.**

If the procedure of the extraction of the theoretical equations regarding carrier number fluctuation effect takes place without considering residual charge, this can lead to very significant conclusions regarding the effect of residual charge on noise behavior. If residual charge is considered insignificant, then it must be eliminated in eqns (A6, A10). This results in the extraction of the following equation regarding 1/f noise due to carrier number fluctuation effect if the equivalent integral of eqn (A8) is solved:

$$\frac{S_{I_D}}{I_D^2} f \bigg|_{\Delta N} = \frac{SD\big|_{\Delta N} \cdot KD\big|_{\Delta N}}{\big[g(V_c)\big]_{V_{GS}^{V_{ol}}}}, SD\big|_{\Delta N} = \frac{2KT\lambda N_T e^2}{(C_{top} + C_{back})kWL} \qquad (Eq.A16)$$

and $KD/_{\Delta N}$ is now given as:

$$\text{for } V_{cs,d} > 0, \quad KD\Big|_{\Delta N} = \Big[ 2\ln\big(C_{top} + C_{back} + kV_c\big) - 4\ln\big(C_{top} + C_{back}\big)\Big]_{V_{cd}}^{V_{cs}}$$

$$\text{for } V_{cs,d} < 0, \quad KD\Big|_{\Delta N} = \Big[ -2\ln\big(C_{top} + C_{back} - kV_c\big)\Big]_{V_{cd}}^{V_{cs}} \qquad \text{(Eq.A17)}$$

eqn (A17) is much simpler that eqn (2) of the manuscript. Regarding Hooge model, residual charge plays a role only in $g(V_c)$ factor in eqn (A10). As it can be seen in Fig. 3a of the manuscript, the omission of the residual charge lead to a $\Lambda - shape$ behavior even for the carrier number fluctuation effect while the less the residual charge, the steeper $\Lambda - shape$ trend with a higher maximum is observed for both carrier number and mobility fluctuation effects.

It would be very useful to observe how the absence of the residual charge affects both noise mechanisms $\Delta N$ and $\Delta \mu$ locally in the transistor channel. Regarding $\Delta N$ local noise model described by eqn (4) of the manuscript and $\Delta \mu$ local noise model described by eqn (A12), residual charge has an effect only in $Q_{gr}$ as this is defined in eqn (A6). As it can be seen in Fig. S2, residual charge does not affect local noise at higher gate voltages for both noise mechanisms as it was expected (see Fig. 3a of the manuscript) since there $\rho_0$ does not affect significantly $Q_{gr}$. On the contrary at CNP, where $\rho_0$ approximately dominates $Q_{gr}$, the effect on local noise mechanisms is important. Fig. S2a shows the increase of $\Delta N$ local noise when $\rho_0$ is ignored leading to the $\Lambda$-shape of Fig. 3a of the manuscript. Similarly Fig. S2b shows the increase of $\Delta \mu$ local noise when $\rho_0$ is ignored.

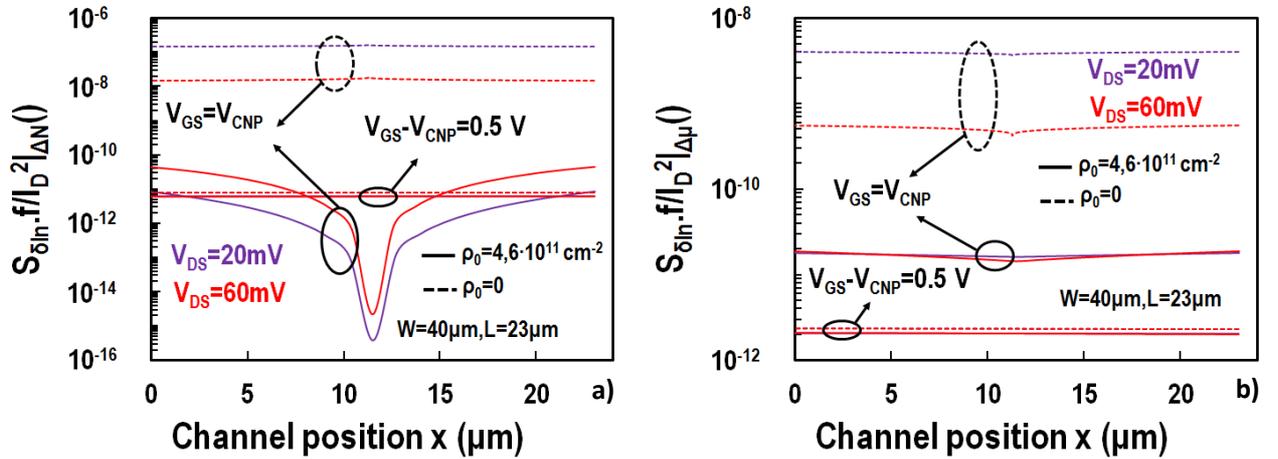

Figure S2. Normalized PSD of the local noise, $S_{\delta in}/I_D^2$, referred to *1 Hz*, vs. channel potential *x* for $\Delta N$ (a) and $\Delta \mu$ (b) noise mechanisms.

**D. Supplementary Information Figure S3: similar analysis with Fig. 4a and 4b of the manuscript but for the rest of the channel lengths**

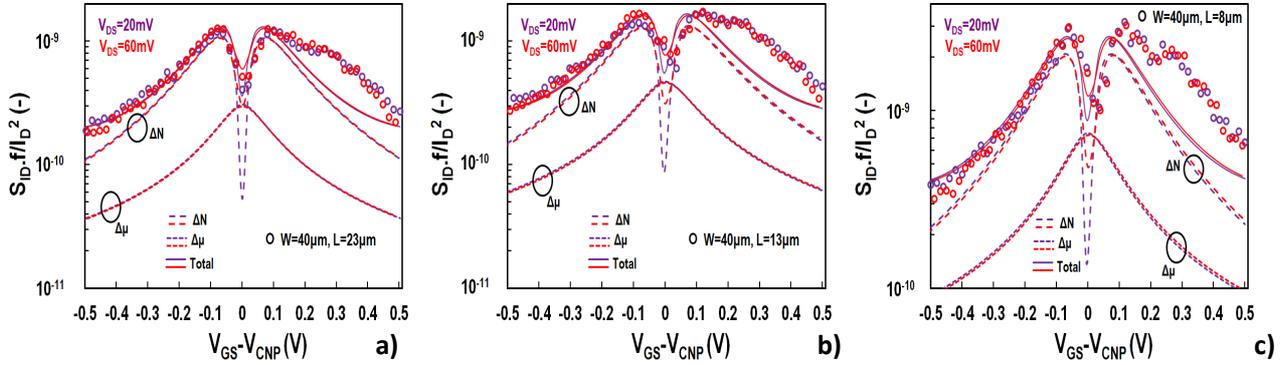

Figure S3. Output noise divided by squared drain current $S_{ID}/I_D^2$, referred to *1 Hz*, vs. top gate voltage overdrive $V_{GS} - V_{CNP}$, for liquid top-gated GFETs with *W=40 μm* for channel length *L=23 μm* (a), *L=13 μm* (b) and *L=8 μm* (c) at $V_{DS} = 20, 60 mV$. markers: measured, solid lines: model, dashed lines: different noise contributions.

**E. Supplementary Information: Derivation of an $(g_m/I_D)^2$ related LFN model with and without correlated mobility fluctuations**

A very common approximation for modeling LFN in Si MOSFETs relates the output noise divided by squared drain current $S_{ID}/I_D^2$, with the squared transconductance to current ratio $(g_m/I_D)^2$ [15-16]. Despite the fact that this model is widely used in circuit simulators, is valid only under uniform channel conditions. This method has also been applied in Graphene FETs [4] and has been found to underestimate LFN at CNP where the channel is non-uniform even for a small $V_{DS}$ as shown in Figure 2c of the main manuscript. In this section we will follow a similar approach as in References 15-17 in order to show how this model is extracted for Graphene FETs with and without the effect of correlated mobility fluctuations. For reasons of simplicity and since back gate voltage is not active in the devices used in this work, both back gate voltage and capacitance will be ignored.

Initially, we will show that the model proposed in Reference 16 ($S_{ID}/I_D^2 = (g_m/I_D)^2 . S_{Vfb}$ ) can be also applied in SLG FETs. From basic GFET electrostatics and if back gate is ignored we have:

$$Q_{gr}(x) = -C_t (V_{GS} - V_{GS0} + V_c(x))$$  (Eq. A18)

From Drift-Diffusion theory [43-45], we can assume that:

$$\Delta I_D = \frac{\mu W}{L} \int_{V_S}^{V_D} \Delta Q_{gr}(x) dV = \frac{\mu W}{L} \int_0^L \frac{dQ_{gr}(x)}{dV_{GS}} \frac{dV_{GS}}{dQ_t} \Delta Q_t \frac{dV}{dx} dx \qquad \text{(Eq. A19)}$$

From eqn (A18) we can conclude that $dV_{GS}/dQ_{gr}(x) = -1/C_t$ while if we assume that $KV_c >> qT$ which means that we are away from CNP and thus $C_q >> C_t$ then from eqns (1, 3) of the main manuscript we have $dQ_{gr}(x)/dQ_t = 1$. So eqn (A19) becomes:

$$\Delta I_D = \frac{-\mu W}{L} \int_0^L \frac{dQ_{gr}(x)}{dV_{GS}} \frac{1}{C_t} \Delta Q_t \frac{dV}{dx} dx = \frac{-\mu W}{L} \int_0^L \frac{dQ_{gr}(x)}{dI_{DS}} \frac{dI_D}{dV_{GS}} \frac{1}{C_t} \Delta Q_t \frac{dV}{dx} dx \qquad \text{(Eq. A20)}$$

Again from Drift-Diffusion theory we have:

$$I_D = \mu W Q_{gr}(x) \frac{dV}{dx} \Rightarrow Q_{gr}(x) = \frac{-I_D}{\mu W \dfrac{dV}{dx}} \qquad \text{(Eq. A21)}$$

Under the assumption of a uniform channel where the graphene charge $Q_{gr}$ and the electric field $dV/dx$ are constant along it we have:

$$\frac{dQ_{gr}}{dI_D} = \frac{-1}{\mu W \dfrac{dV}{dx}} \qquad \text{(Eq. A22)}$$

From eqns (A20, A22) and since $d_{ID}/dV_{GS} = g_m$ we conclude:

$$\Delta I_D = \frac{1}{LC_t} g_m . e \int_0^L \Delta N_t dx \qquad \text{(Eq. A23)}$$

Which leads to:

$$\frac{S_{I_D}}{I_D^2} f \bigg|_{\Delta N} = \left(\frac{g_m}{I_D}\right)^2 . \frac{q^2 KT \lambda N_T}{WLC_t^2} \qquad \text{(Eq. A24)}$$

The above eqn is exactly the same with eqn (9) of Reference 16 with the constant term to represent the flat band voltage fluctuations $S_{VFb}$. As we proved before, this model is valid only under uniform channel conditions and away from the CNP.

According to Reference 17, the model of eqn (A24) can be expanded including the correlated mobility fluctuations as:

$$\frac{S_{I_D}}{I_D^2} f \bigg|_{\Delta N} = \left(\frac{g_m}{I_D} + \alpha_c \mu C_t\right)^2 . \frac{q^2 KT \lambda N_T}{WLC_t^2} \qquad \text{(Eq. A25)}$$

where $\alpha_c$ is the Coulomb scattering coefficient in *V.s/C* and $\mu$ is the mobility of the device. Figure S4 below presents the behavior of this simple approach described above with (eqn A25) and without (eqn A24) the effect of correlated mobility fluctuations for the shortest device with *L=5.5 μm* at *$V_{DS}$=20 mV* and *$V_{DS}$=60 mV*.

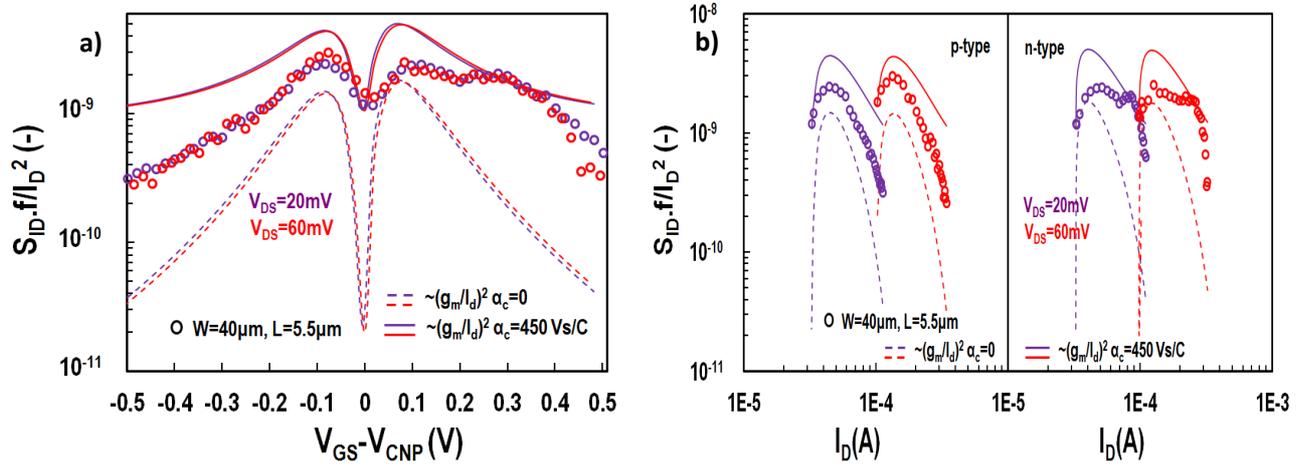

Figure S4. Output noise normalized with area and divided by squared drain current *$S_{ID}/I_D^2$*, referred to *1 Hz,* vs. top gate voltage overdrive *$V_{GS} - V_{CNP}$* (a) and vs. drain current in both p- and n-type region (b) for liquid top-gated GFETs with *W/ L=40 μm / 5.5 μm* at *$V_{DS}$=20 mV* and *$V_{DS}$=60 mV*. markers: measured, solid lines: eqn (A25) model, dashed lines: eqn (A24) model.

Figure S4a presents the normalized *$S_{ID}/I_D^2$* LFN vs top gate voltage overdrive *$V_{GS} - V_{CNP}$* and what can be observed is that the model of eqn (A24) (*$\alpha_c=0$*) underestimates LFN as it is also shown in Figure 4a of the manuscript for the longest device. Furthermore it is clear that the behavior of LFN is independent of *$V_{DS}$* even at the CNP because of the consideration of a uniform channel. If correlated mobility fluctuations model of eqn (A25) is activated then for a value of *$\alpha_c=450$ Vs/C* the model captures the level of LFN at CNP still with no drain voltage dependence. But simultaneously the model overestimates LFN at higher gate voltages. Even if we assume that with an appropriate combination of *$\alpha_c$* and *$\alpha_H$* parameters we could achieve a better fitting, still the model would be independent of *$V_{DS}$* due to the homogeneous channel consideration.

Figure S4b presents the results of Figure 4a versus drain current *$I_D$* in log scale. Since *$I_D$* is symmetrical below (p-type) and above (n-type) CNP as it is shown in Figure 1c of the main manuscript, the two regions should be shown separately in log-scale. In an illustration similar to Figure S4b for Si MOSFETs, *$S_{ID}/I_D^2$* LFN is maximum and constant in weak inversion region and decreases as we get deeper in strong inversion

(See Figure 6 of Reference 17). Regarding weak inversion regime, this occurs because $g_m/I_D$ term is maximum and constant in the specific region and thus, eqn (A24) becomes equivalent to eqn (A25) since $\alpha_c$ is negligible. Consequently, $N_T$ parameter which is included in $S_{Vfb}$ term is extracted. As the drain current gets higher, LFN decreases and $\alpha_c$ parameter is extracted from this higher current regime. This is not the case in GFET though as it can be seen from Figure S4b since $(g_m/I_D)^2$ is not constant in lower current regime.

**F. Supplementary Information Figure S5: normalized output noise with device area - $WLS_{ID}/I_D{}^2$**

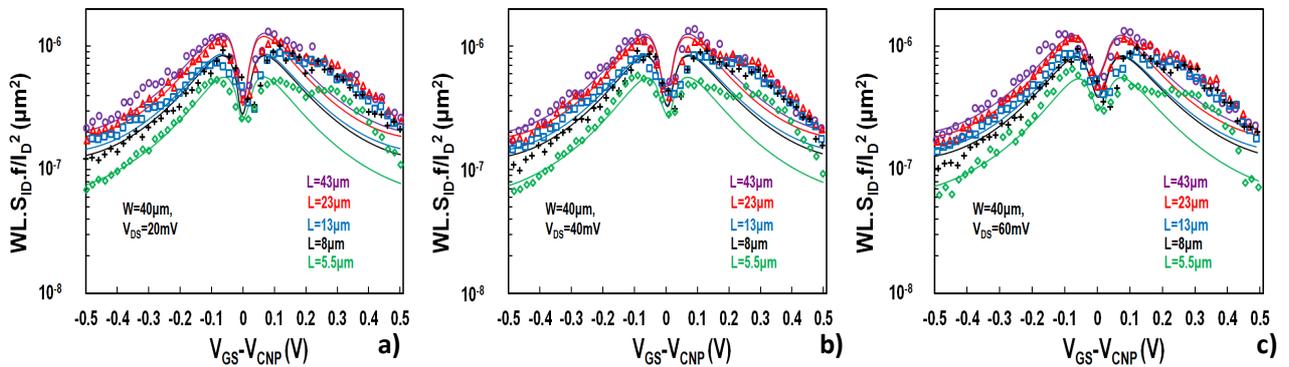

Figure S5. Output noise normalized with area and divided by squared drain current $WLS_{ID}/I_D{}^2$, referred to *1 Hz*, vs. top gate voltage overdrive $V_{GS} - V_{CNP}$ for liquid top-gated GFETs with *W=40 μm* for different channel length values (*L=43, 23, 13, 8, 5.5 μm*) at *$V_{DS}$=20 mV* (a), *$V_{DS}$=40 mV* (b) and *$V_{DS}$=60 mV* (c). markers: measured, solid lines: model.